\documentclass[useAMS]{mn2e}
\usepackage{times}
\usepackage{psfig}
\usepackage{epsfig}
\usepackage{graphicx}

\def\lsim{\lower.5ex\hbox{$\; \buildrel < \over \sim \;$}}
\def\gsim{\lower.5ex\hbox{$\; \buildrel > \over \sim \;$}}

\def\lsim{\lower.5ex\hbox{$\; \buildrel < \over \sim \;$}}
\def\gsim{\lower.5ex\hbox{$\; \buildrel > \over \sim \;$}}

\def\be{\begin{equation}}

\def\ee{\end{equation}}

\begin{document}

\title[An origin for the main pulsation and overtones of SGR1900+14...]{An 
origin for the main pulsation and overtones of SGR1900+14 during the august 
27 (1998) superoutburst}

\author[Mosquera Cuesta]{Herman J. Mosquera Cuesta$^{1,2,3}$ \\
$^1$ Abdus Salam International Centre for Theoretical Physics,
Strada Costiera 11, Miramare 34014, Trieste, Italy \\ $^2$ Centro
Brasileiro de Pesquisas F\'{\i}sicas, Laborat\'orio de Cosmologia e
F\'{\i}sica Experimental de Altas Energias \\ \hskip 0.2 truecm 
Rua Dr. Xavier Sigaud 150,
Cep 22290-180, Urca, Rio de Janeiro, RJ, Brazil \\ $^3$ Centro
Latino-Americano de F\'{\i}sica, Avenida Wenceslau Braz 173, Cep
22290-140 Fundos, Botafogo, Rio de Janeiro, RJ, Brazil }

\maketitle

\date{\today}

\begin{abstract}
The crucial observation on the occurrence of subpulses (overtones) in
the Power Spectral Density  of the August 27 (1998) event from
SGR1900+14, as discovered by BeppoSAX (Feroci et al. 1999), has
received no consistent explanation in the context of the competing
theories to explain the SGRs phenomenology: the magnetar and
accretion-driven models. Based on the ultra-relativistic, ultracompact
X-ray binary model introduced in the accompanying paper (Mosquera
Cuesta 2004a), I present here a self-consistent explanation for such an
striking feature. I suggest that both the fundamental mode and the
overtones observed in SGR1900+14 stem from pulsations of a massive
white dwarf (WD). The fundamental mode (and likely some of its
harmonics) is excited because of the mutual gravitational interaction
with its orbital companion (a NS, envisioned here as point mass object)
whenever the binary Keplerian orbital frequency is a  multiple integer
number ($m$) of that mode frequency (Pons et al. 2002). Besides, a
large part of the powerful irradiation from the fireball-like explosion
occurring on the NS (after partial accretion of disk material) is
absorbed in different regions of the star driving the excitation of
other multipoles (Podsiadlowski 1991,1995), i.e., the overtones (fluid modes
of higher frequency). Part of this energy is then reemitted into space
from the WD surface or atmosphere.  This way, the WD lightcurve carries
with it the signature of these pulsations inasmuch the way as it
happens with the Sun pulsations in Helioseismology.  It is shown that
our theoretical prediction on the pulsation spectrum agrees quite well
with the one found by BeppoSAX (Feroci et al. 1999). A feature
confirmed by numerical simulations (Montgomery \& Winget 2000).
\end{abstract}



\section{The August 27 (1998) event and magnetar model: Concordance or crisis}

The lightcurve of the spectacular superoutburst from SGR 1900+14 in August 
27, 1998 exhibited a stable pulsation with period 5.16s (Hurley 1999a,b,c;
Murakami et al. 1999; Mazets et al. 1999; Feroci et al. 1999). Since
the modulation frequency is in the range of the other three SGRs
studied before, Kouveliotou et al. (1999) and Hurley et al. (1999a,b,c)
concluded that the observations provide strong support to the Duncan \&
Thompson (1992); Thompson \& Duncan (1995,1996) magnetar model for
SGRs. They claimed that the observed spindown rate of the pulse period,
$\dot{P} = 1.1 \times 10^{-10}$ss$^{-1}$, may be explained by emission
of dipolar radiation from an NS endowed with a very strong magnetic
field $B \sim (2-8) 10^{14}$G, a characteristic magnetic field strength
inferred also from the spin down of the pulse period $P = 7.47$s of SGR
1806-20 (Kouveliotou et al. 1998).

Despite the rough agreement between the SGR1900+14 on August 27, 1998
observations and the theoretical prediction of the magnetar picture,
problems for this model came together with that apparent success. On
the one hand, it is clear that there is an overall consistency between
the observations and the magnetar model. However, it is also clear that
the simple ``giant dipole" dynamics is not unique in explaining the
spindown history of the objects, and in fact several problems with that
picture are already present when dealing with ordinary, low-field
pulsar cousins. Among them we can quote the values of the few measured
braking indexes (which do require modifications from the canonical
dipole spindown model) and the mismatch between characteristic and
dynamical-historical ages in the case of some PSR-SN associations.
Actually, the presence of ordinary pulsars in the high-$P$, high-${\dot
P}$ ``magnetar" region remains puzzling (Manchester 2000), and an
interpretation of the SGR-AXPs in terms of high (but sub-Schwinger)
fields is in principle possible (Allen \& Horvath 2000). On the other
hand, it seems that a single population can not account for both ordinary
and magnetar objects (Regimbau \& de Freitas Pacheco 2001).\footnote{ 
The author truly thanks Prof. J. Horvath (IAG-USP/Brazil) for enlightning discussions on these issues.}

The event GRB980827 was also detected by BeppoSAX (Feroci et al.
1999).  These observations led to the discovery of an extremely regular
interpulse set in the X-ray data of GRB980827 event from SGR 1900+14
(see Fig.1).
The interpulses appear separated in time $\sim 1.1$s in between, with
no lag. This behavior, Feroci et al. (1999) advanced, is unexpected and
quite difficult to explain in the magnetar framework.  According to the
magnetar model for SGRs, global seismic oscillations (Duncan 1998),
pure shear deformation-induced toroidal modes (standardly labelled as
$_lT_n$) with no radial components, i.e., $n = 0$ {\it overtones}, are
expected to be produced in association with the onset of a new
recurrence of a ``soft" $\gamma$-ray repeater. According to Duncan
(1998): ``... these toroidal modes are easy to excite via starquakes
because the restoring force is determined uniquely by the weak Coulomb
forces of the crustal ions.  However, {\it overtones} are not allowed
because that would require far too much energy so as to allow for the
extremely short period ($\leq 1$~ms, or {\it lower}) NS oscillations to
be excited". An energy that the crust cracking mechanism cannot provide
(de Freitas Pacheco 1998).

In overall, the simplest and neatest interpretation of that observations
is that the subpulses in the lightcurve (Power Spectral Density) of the
burst from GRB980827 are {\it overtones} of the fundamental frequency
$f_0 = 0.194$~Hz (Feroci et al. 1999). As shown below, we may be
actually detecting the  whole WD pulsation spectrum, up to 19 harmonics
(Feroci et al. 1999). The importance of having a large part of the
pulsational spectrum of the object can not be overstated. The main
purpose of this Letter is to address this issue. We show that a
self-consistent interpretation for the fundamental mode of pulsation
and the harmonics discovered by Feroci et al. (1999) could be
constructed invoking the excitation of WD pulsational $p-modes$ driven
by tidal-heating (Pons et al. 2002) plus the irradiation (Podsiadlowski 
1991) triggered by supercritical accretion onto its orbital NS partner 
in an X-rays ultracompact ultra-relativistic binary (see a more detailed 
discussion on this model in the accompanying paper by Mosquera Cuesta 
2004a).

\section{The WD excitation energy source: Tidal heating plus 
$\gamma$-irradiation}


The basis of the ultra-relativistic compact binary model for SGRs
(Mosquera Cuesta 2004a), is that during rather sparse catastrophic
epochs ($\Delta T_{\rm SGRs} \leq 10$yr) the WD starts to transfer mass
onto a low-magnetized ($B \sim 10^{10}$G) rapidly rotating millisecond
(ms) massive NS ($\sim$ 2M$_\odot$). This process develops via the
formation of a thick dense massive accretion disk (TDD) very close to
the innermost stable circular orbit around the NS. The disk becomes
unstable due to gravitational runaway or Jeans instability, partially
slumps and inspirals onto the NS. The abrupt supercritical mass
accretion onto the NS releases a quasi-thermal powerful $\gamma$-ray
burst (GRB), a fireball to say. A parcel of the accretion energy
illuminates with hard radiation ($\gamma-rays$) the WD, additionally
perturbing its hydrostatic equilibrium. The WD absorbs this huge energy
at its interior and atmosphere.\footnote{ Podsiadlowski (1991; and
references therein) has discussed this process for irradiated main
sequence and evolved stars, where the companion expands to a new state
of thermal equilibirum, which provides a new mechanism to drive mass
transfer onto the NS. This alters the binary evolution, and may bring a
new evolutionary stage during which the orbital period increases,
leading to larger orbital period during and at the end of the mass
transfer. Although the WD in this relativistic binary is a degenerate
compact star with an atmosphere, we believe a similar behavior should
also take place in it, since the new thermal timescale, $\propto \Delta
R_{\rm WD}$, is comparable to the heat ($\gamma$-rays) difussion time
in the outer layers.} This irradiation excites other $p$-modes of the
WD oscillation spectrum, the {\it overtones}, since it already pulsates
at its fundamental mode because of the tidal interaction of the binary
(see Table I).

\begin{table}
\centering
\begin{minipage}{80mm}
\caption{First few radial (l=0) and nonradial (l$\neq$0) modes for a 1.05
~M$_\odot$ WD model (Temperature =12000 K). Columns 1, 2, 4 and 5, represent 
the l-value, radial overtone value n, period (s) and frequency (Hz), as 
computed by Montgomery \& Winget (1999).\label{tbl-1} } 
\begin{tabular}{ccccc}
\hline
l & n & { } & P [s] & {Freq. [Hz] }  \\
\hline
0  & 1 &   .66121  &  5.416  &  .1846   \\
 0  & 2 &  1.87364  &  1.911  &  .5232  \\
 0  & 3  &  2.80941 &  1.275 & 7 .7845  \\
 0  & 4  & 3.69229  &   .970  & 1.0310  \\
0 &  5  & 4.54111   &  .789 &  1.2680   \\ 
 1 &  1  & 1.39589  &  2.566  &  .3898  \\
 1 &  2 &  2.33949  &  1.531  &  .6532  \\
 1 & 3  & 3.23226  &  1.108  &  .9025   \\
 1 &  4 &  4.09340  &   .875 &  1.1430  \\ 
 2  &  0  &  .99360  &  3.604  &  .2774  \\
 2  & 1 &  1.85472  &  1.931   & .5179  \\
 2  & 2 &  2.73775  &  1.308  & .7644  \\
 2  & 3  & 3.60567   &  .993  & 1.0068  \\
 2  & 4 &  4.45214   &  .804  & 1.2431 \\
 3 &  0 &  1.24455  &  2.878  &  .3475  \\
 3  & 1  & 2.17168 &   1.649  &  .6064  \\
 3 &  2  & 3.06016  &  1.170  &  .8545  \\
 3  & 3  & 3.92616   &  .912  & 1.0963 \\
 3 &  4  & 4.76880  &   .751 &  1.3316  \\
\hline
\end{tabular}
\end{minipage}
\end{table}


The abrupt supercritical accretion also perturbs the NS hydrostatic
equilibrium which drives it into non-radial nonaxisymmetric
oscillations that produces GWs due to excitation of the NS fluid modes
(see Mosquera Cuesta et al. 1998). Further, the remaining part of the
TDD might be the neighbour environment  ($\sim 100$km from the NS)
where matter carried by the fireball can nucleosinthesize to produce
the noticeable iron Fe$^{56}$ line discovered by Strohmaier \&  Ibrahim
(2000) during the GRB980827 giant outburst. This possibility is
explored in a forthcoming communication (Mosquera Cuesta, Duarte \& de 
Freitas Pacheco, in preparation). (We address the reader to the related 
paper by Mosquera Cuesta 2004a, for a full description of the interacting 
relativistic compact binary here pictured).

\subsection{A possible origin for the fundamental mode frequency and  
overtones}


In this section we argue that the theoretical modeling of WD pulsations
as performed by Montgomery \& Winget (1999) can accurately account for
the spectrum of frequencies discovered by BeppoSAX during the event
GRB980827 from SGR1900+14 (Feroci et al. 1999). We note in passing that
although the Montgomery \& Winget (1999) numerical simulations focussed
on the study of long period gravity-induced WD oscillations, which are
highly relevant for the potential identification by the Whole Earth
Telescope of pulsating ZZ-Ceti WDs in surveys, those models also
provide the spectrum of pulsations of the short period pressure-driven
({\it p-mode}) pulsations.

Any perturbation of the hydrostatic equilibrium of a canonical
WD will grow on its dynamical timescale 

\be
\tau_{dyn} \sim (G \rho_{{WD}})^{-1/2} \sim 5.16~{\rm s} 
\left(\frac{4\times 10^6~{\rm g~cm^{-3}} } {\rho_{\rm WD}}\right)^{1/2}\; .
\ee

The WD normal mode spectrum ($p-modes$) is obtained from the radial 
wavenumber $k_r$ defined by (Montgomery \& Winget 1999) 

\be
k_r^2 = ({1/\sigma^2 c_s^2}) (\sigma^2 - L_l^2)(\sigma^2 - N^2)
\ee

with $\sigma$ the mode angular frequency and $c_s$ the sound speed in the 
star material. Here the squared Lamb/acoustic frequency is defined

\be
L_l^2 = l(l+1) \left[\frac{c_s^2}{ r^2}\right]
\ee

where $r$ is the radial variable, and $N^2$ the Brunt-V\"ais\"al\"a frequency. 
For $p-modes$:

\be
\sigma^2 > L^2_l, \hskip 0.2 truecm N^2,
\ee

and 

\be
\sigma \sim   \frac{ {k_r}{\pi} } { \int^{r_2}_{r_1} { \frac{dr}{c_s} } },
\ee

where $r_2$, $r_1$ are the inner and outer turning points,
respectively, at which $k_r = 0$ for a given $\sigma$.

Montgomery \& Winget (1999) studied the pulsational modes of a massive
WD ($M_{\rm WD} \sim 1.1 {\rm M}_\odot$) which possess a hydrogen
atmosphere of about M${^{crust}_{\rm WD}} \sim 10^{-6} {\rm M}_\odot$, with
and without core-crystallization. The full spectrum of one of their
numerical models is displayed in Table I, and a selected subset is
displayed in Table II\footnote{A curious note on this result is that it
exhibits the same number of overtones as the one Feroci et al. (1999)
found in the GRB980827 data from SGR1900+14.} to compare with BeppoSAX
observations (Feroci et al. 1999).

\begin{table}
\centering
\begin{minipage}{80mm}
\caption{A subset of the WD frequencies displayed in Table I, versus the 
modulation  spectrum from SGR 1900+14 on August 27 (1998) as discovered by 
Feroci et al. (1999). \label{tbl-2} }
\begin{tabular}{cccc}
\hline
{Overtones.} &  {Num. Model [Hz]} & {BeppoSAX [Hz]} & {Mismatch (\%)} \\
\hline
{l= 0, n = 1} & { 0.1846} & {0.194} & { 5.0} \\
{l= 1, n = 1} & { 0.3898} & {0.389} & {0.2}  \\
{l= 2,0, n = 2,3} & {0.7644, 0.7845} & { 0.775} & {1.0, 1.23} \\  
{l= 1,2, n = 3,3} & {0.9025, 1.0068} & {0.969} & {4.54, 3.78} \\
{l= 1, n = 4} & {1.1430} & { 1.161 } & {1.55} \\
\hline
\end{tabular}
\end{minipage}
\end{table}


\section{Discussions}

\subsection{The WD required mass}

 If we use the picture being introduced here to explain the pulsation
 discovered in GRB980827 (Hurley 1999a; Kouveliotou 1999a; Murakami et
al. 1999; Feroci et al. 1999), it is easy to see that the WD mass (see
Table 3) needed to produce pulsations with timescale of 5.16s is
$\sim 1.1$~M$_\odot$ (Montgomery \& Winget 1999; Shapiro \& Teukolsky
1983).  The interpulses discovered by the Italian team of the BeppoSAX
collaboration can also be explained in a simple manner (except that
there are instrumental errors): they are a kind of WD {\it ringing
overtones}. The full attenuation of the oscillations will be achieved
on a much longer timescale via the {\it wave-leakage of radial modes}
as discussed by Hansen, Winget \& Kawaler (1985), see below.  So the
observed pulsations may stem from a WD, not from a slowly rotating
hypermagnetised NS.

\begin{figure}
{\centering \hskip 2.0 truecm \epsfxsize=6.6cm \epsfbox{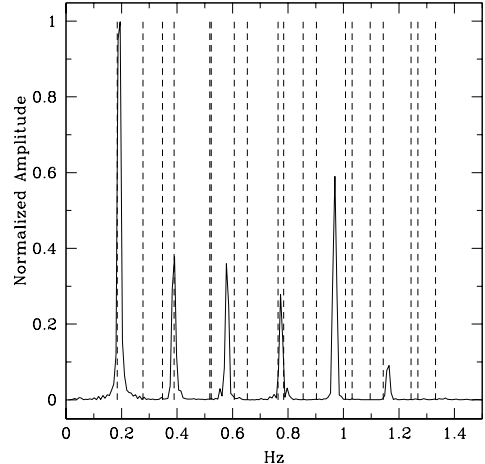}  }
\caption{Graphic comparison between the observed modulation (main six overtones) 
in SGR1900+14 as detected by BeppoSAX (Feroci et al. 1999) and the theoretical pulsation spectrum (vertical dashed lines) computed by Montgomery \& Winget 
(1999). We claim that the overtone of frequency near 0.96~Hz might have been 
enhanced in power via irradiation during the $\gamma$-rays superoutburst.}
\end{figure}

\subsection{SGRs spindown rate}

Moreover, the SGR 1900+14 spindown determined by Kouvelioutou et al.
(1999) and the pulse period increase found by Murakami et al. (1999);
Marsden, Rothschild \& Lingenfelter (1999); Harding, Contopoulos \&
Kazanas (1999) may be explained as follows: because the WD is a
gravothermal system, as soon as it loses mass (when overflowing its
Roche lobe) its negative specific heat forces it to expand until a new
dynamical equilibrium radius is found.  Then, the next stage of
pulsation should occur with a slightly longer period compared to the
previous one (secular increase).  As a result, the pulse period
increases. However, the interplay between the WD thermal cooling during
the post-outburst reemission (which implies post-burst shrinking) and
tidal heating via gravitational interaction with the NS (which implies
pre-burst expansion) may lead to a slight up-down-up change in the
fundamental mode of pulsation.  This processes may help to understand
the observed tiny changes in the SGR 1900+14 period: near superoutburst 
it is a bit longer, while in quiescence it shortens.

\begin{table}
\centering
\begin{minipage}{60mm}
\caption{SGRs observed pulsation periods and inferred masses for 
the WD in the binary model. \label{tbl-3} }
\begin{tabular}{ccc}
\hline
{ SGR } & {P [s] } & {Mass [M$_\odot$] } \\
\hline
{1900+14} & 5.16 &  { 1.1 }  \\ 
{0526-66} & 8.1 & { 0.70 } \\ 
{1806-20} & 7.47 & 0.80 \\ 
{1627-41} & 6.7 & {0.95 } \\
\hline
\end{tabular}
\end{minipage}
\end{table}

Thus, the rate of variation of the viscously attenuated pulsation
time, $\tau_{pulse} = 5.16$~s, divided by the timescale $\tau_{w-l} \sim
10^3$~yr needed to dissipate via {\it wave-leakage} for {\it radial
modes}\footnote{Physical support for the figures that have been used here
comes from the detailed numerical calculations of pulsation periods for
DA and DB WDs given by Hansen, Winget \& Kawaler (1985), where
timescales for dissipation of energies such as the one absorbed by the
WD in this scenario are estimated for wave-leakage via unstable radial
modes with $e$-folding timescale $\tau_D \sim 10^3$~yr, for nonadiabatic
driving of the mode with energy $10^{44}$~erg.} the absorbed energy
$\sim 10^{44}$~erg, gives us the observed ``mean" spindown rate in SGR 
1900+14 (here we assume this the dominant source available to the WD)

\be
\left(\frac{\partial{\tau_{pulse}} }{ \partial{\tau_{w-l} }}
\right) \equiv \frac{ 5.16 \; {\rm s}}{3.17 \times 10^{10}\; {\rm s} } =
1.63\times 10^{-10} \; {\rm s} {\rm s}^{-1}\; .
\ee

In our view this is the origin of those important time variations.
Discrepancies might be caused by the differences in the WD model used
(see Table 1). We stress that in a realistic star the overtone
frequencies depend on the WD matter EOS and the degree of core
crystallization.  Changes in pulsation (fundamental) periods of $\sim $
a few percent are expected to occur once lattice crystallization onsets
while the harmonics (overtones) remain almost unchanged (Montgomery \&
Winget 1999).

\subsection{X-ray emission during quiescent, pre- and post-burst }

What about  X-ray emission and pulsation in long-term quiescence, pre-
and post-burst? A number of viable processes may explain why SGRs glow in
hard X-rays over some months before and after undergoing dramatic
transients such as GRB980827 in SGR 1900+14, as well as the long-term
quiescent emission:  1) changes in the gravitational potential at each
orbital revolution (Podsiadlowski 1991,1995), 2) crust cracking driven by the
WD strong magnetic field  (analogously to the NS case, see de Freitas
Pacheco 1998), and 3) plate tectonics (Rothschild, Marsden \&
Lingenfelter 2001) induced by any or both the mechanisms just listed.

In this paper, as a first approach, we shall focus our discussion on
this issue in the context of the scenario number 2) above.  The
remaining possibilities will be addressed elsewhere.  This key property
can be explained if we suppose that the quiescent soft X-rays
luminosity $L^{SGRs}_{X} \sim 10^{35}$~erg~s$^{-1}$ (Kouveliotou et al.
1999) is powered by the release of WD crustal elastic
energy\footnote{An original suggestion by Prof.  Posiadlowski (2000),
Dept. of Physics, Oxford University, England, UK.}, inasmuch as in the
current picture of magnetars (de Freitas Pacheco 1998). In this case,
we get

\be
\frac{B_{surf} B_{core}}{8 \pi \mu_0} \times 4\pi R^2_{WD} \times \Delta 
R_{WD}\geq L{^{SGRs}_X} \times \tau
\ee

where  $\tau \sim 10^4$~yr is the system lifetime since its formation,
$B_{surf} \sim 10^9$~G WD crustal magnetic field, as in the super
magnetised WD: RE J0317-853, $M_\odot = 1.35$, P = 725~s
(Wickramasinghe \& Ferrario 2000), and $B_{core} \sim 10^{13}$~G is the
WD core magnetic field. This yields a WD crust  thickness: $\Delta
R_{WD} \sim$ 70~km\footnote{Note that this thickness is nearly equal to
the Earth's {\it astenosphere} height scale. These scaling similarities
(including the WD radius) may suggest a plate tectonics strikingly
similar to that observed on our planet. Some suggestions in this
direction have been put forward by Cheng et al.  (1995), but they were
associated to NS plate tectonics rather than to a WD, which in many
respects is alike our planet.}.  The crust-cracking induces shear
stresses which dissipate energy causing excitation of WD oscillation
modes quite similiar to the NS case (de Freitas Pacheco 1998, and
references therein) \footnote{We add to this mechanism that is possible
for the (ms)NS rotation energy: $E_{spin} \sim I_{NS} \Omega{^2_{NS}}
\sim 3 \times 10^{35}$~erg~s$^{-1}$ to play some role in driving the
X-rays emission and pulsation during these stages due to its
irradiation onto the WD, although the tidal interaction must be the
dominant source.}.  We thus emphasize that the conclusion by Mazets et al.
(1999):  ``... the processes accounting for emission of the narrow
initial pulse and the long pulsating tail in both SGR 0526-66 and
1900+14 {\it are separated in the source not only on time but in
space...''}, is realized in the picture introduced here, i. e., the NS
releases the superoutburst while the WD the subsequent tail of
pulsations via the mechanism invoked above.

\section{Conclusions}

As a summary, the NS low magnetization in this model leads SGR1900+14
(and perhaps all SGRs) under the pulsar death line making it
undetectable as a binary radio pulsar, a viewpoint confirmed by
Xilouris et al. (1998).  Overall, since several SGRs are enshrouded by
intervening galactic dust and gaseous nebulae (SNe remnants: Gaensler
et al. 2001), optical observations of the suggested hot WD ($\sim
12000$~K) may render a breakthrough.  This fact makes it a systematic
high resolution search for such optical (or infra-red) counterparts of
SGRs a timely and promissing task. In this line, we remark that
Ackerlof et al. (2000) have pursued optical follow-ups of SGRs over 10
outbursts: 8 events from SGR1900+14 and 2 events from SGR 1806-20.
Although a careful search for new or variable sources in the view field
of those SGRs was performed, no optical counterparts were seen down to
$m_V \sim 15$. Thus the search for such SGRs counterparts remains, and
perhaps VLA, GEMINI or KECK  observations in the K-band could
unravel them.  On the other hand, it will be extremely relevant if the
{\it propeller model} of Marsden, Rothschild \& Lingenfelter (2001) and
Alpar (1999, 2000) could provide a clean explanation for the observed
spectrum of pulsational modes.  This may help to discriminate among the
SGRs accretion models through future observations.

\thanks{We deeply thank Mike Montgomery and Don Winget for kindly
running their numerical code to simulate WD non-radial {\it p-mode}
pulsations on our request, and also for the many fruitful discussion 
on this subject.}

\end{document}